\documentclass[11pt,article]{amsart}
\usepackage{amsmath}
\usepackage{amssymb}
\usepackage{verbatim}
\usepackage[pdftex]{graphicx}
\usepackage{sidecap}
\usepackage{float}

\theoremstyle{definition}

\theoremstyle{remark}

\numberwithin{equation}{section}

\newcommand{\E}{\mathbb{E}}
\renewcommand{\P}{\mathbb{P}}

\newcommand{\Q}{\mathbb{Q}}

\newcommand{\sign}{\mathrm{sign}}

\newcommand{\abs}[1]{\lvert#1\rvert}

\begin{document}

\title[Volatility smile and out-of-the-money calls]{On volatility smile and an investment strategy with out-of-the-money calls}

\begin{abstract}
A motivating question in this paper is whether a sensible investment strategy may systematically contain long positions in out-of-the-money European calls with short expiry. 
Here we consider a very simple trading strategy for calls. The main points of this note are the following. First, the presented trading strategy appears very lucrative
in the Black-Scholes-Merton (BSM) framework. In fact, it is such even to the extent that the BSM model turns out to be, in a sense, incompatible with the CAPM. Second, if one wishes to 
adapt these models together, then the adjustment of the consistent pricing rule (i.e. modifying state price densities) inevitably leads to some 
form of volatility smile and this is the main point of the paper. Moreover, these observations arise from purely structural considerations. 
\end{abstract}


\author{Jarno Talponen}
\address{University of Eastern Finland, Department of Physics and Mathematics, Box 111, FI-80101 Joensuu, Finland}
\email{talponen@iki.fi}
\date{\today}
\maketitle


\section{Introduction}

In this note we make some simple observations about the prices of European call options in the Black-Scholes-Merton (BSM)
model. We suspect that many scholars and practitioners find the conclusion somewhat 
counter-intuitive\footnote{For example in \cite{Haug} pp. 54-65 the compatibility of CAPM and BSM models is discussed in several occurrences.}. The main observation of this note is that in the confinements of the BSM model it is in some sense 
a very lucrative strategy to keep buying cheap out-of-the-money calls with \emph{short} 
time horizons in successive instances. In fact, the strategy appears to be 'too good to be true'. 
We will work inside the BSM framework with hardly any empirical information, so the findings are \emph{structural} 
in nature.     

The author came across the key observation here by an (unsuccessful) attempt to apply the BSM framework in pricing 
catastrophe bonds. Namely, it would appear like a sound approach to model a CAT bond by means of digital options, 
which are triggered by an extreme behavior of the underlying, or a `tail event'. 
The author tried to find a kind of asymptotic price of risk for an extremely rare but expensive event. The idea was 
to analyze European calls, with eventually extreme parameters, in such a way that the probability of the 
the portfolio yields being non-zero would diminish but the expected value of the yields would be kept constant, say $1$.
This approach somewhat resembles 'renormalization' techniques in physics. 

Here is what went wrong. Suppose that $\P$ and $\Q$ are the physical and the risk-neutral probability measures, respectively, appearing in the standard BSM model. 
Let us recall that $\P$ provides the modeled real probabilities of events, whereas $\Q$ is applied in calculating the prices of derivatives. We will later recall how to calculate 
the Radon-Nikodym derivative $\frac{d\Q}{d\P}$ (a.k.a. stochastic discount factor) for a given equity value $S_T$ 
at the time horizon $T$. The following simple observation is promoted here; namely that the measures $\P$ and $\Q$, although equivalent 
in the measure theoretic sense\footnote{Perhaps the most common name for $\Q$ is the equivalent martingale measure.}, are in a sense \emph{not} well \emph{comparable} financially. 
That is, the ratio $\frac{d\Q}{d\P}|_{S_{T}=K}$ tends to $0$ as $K\to \infty$ and tends to $\infty$ as $K\to 0$. 
Therefore the above mentioned attempt in pricing CAT bonds was bound to fail, since selecting either of the tails results 
in a trivial asymptotic price of risk, either $0$ or $\infty$, depending on which tail is chosen. 

The fact that the above ratio strongly biases small values of $S_T$ is intuitive from the point of view of 
a risk averse agent. Still, it enables one to device an interesting trading strategy. The key idea in forming 
this strategy is that according to the asymptotics of the ratio $\frac{d\Q}{d\P}|_{S_{T}=K}$ it is possible to buy 
'lottery tickets' at a price, which is very small compared to the expected payoff. 

As the result, we will device a rather simple trading strategy which shares some characteristics 
of high-frequency trading strategies, such as a large number of relatively small, restricted bets and 
a very high Sharpe ratio.

One theoretically intriguing phenomenon here is that the outcome of the trading strategy is not well in harmony 
with Modern Portfolio Theory (MPT) or Capital Asset Pricing Model (CAPM) and it produces unrealistically high Sharpe ratios, as already mentioned. It is also quite 
easy to see how moderately out-of-the-money calls have a low beta with respect to the underlying asset. The compatibility 
question of the BSM and CAPM has been addressed in previous literature, see e.g. \cite{AviBick}, \cite{Benninga}.
This is of course a natural problem since BSM and CAPM are typical representatives of the two most commonly applied financial valuation schemata, risk-neutral (RN) pricing and equilibrium pricing, respectively.  Also, Black and Scholes applied CAPM in their seminal paper to give an alternative derivation for their 
option valuation formula. This, per se, does not guarantee the general compatibility of these pricing frameworks. 
It is also well-known that the actuarial value of derivatives may easily differ from the risk-neutral pricing based value.\footnote{Actually, starting from the fact that the treatment of risk premia 
sometimes varies, the actuarial value is not even completely uniquely defined throughout the literature.} 

To reiterate, it turns out that the investigated investment strategy, taken to the extreme, produces in the BSM framework Sharpe ratios tending to the infinity, a state non-maintainable in view of the philosophy of the Modern Portfolio Theory (MPT), so that apparently the BSM model and the MPT are not fully compatible. The same conclusion is valid for BSM and CAPM, 
since the strategy simultaneously produces low betas as well.

We conclude by discussing an interesting connection to volatility smile. Our findings suggest how some form of volatility 
smile can be actually an \emph{expected} phenomenon by looking at structural properties of common valuation models with 
rather minimal empirical information; we mainly rely on the fact that the market price of risk is positive.
In particular, we are \emph{not} required to invoke further asset dynamics considerations (e.g. fat tails), empirical or stylized facts, preference structure considerations, behavioral finance issues, etc. 
The volatility smile here is of formal nature and we are by no means claiming that it fully explains the one witnessed empirically.

We have tried to make the discussion accessible to an audience as general as possible.

\subsection{Preliminaries}

We will make rather casually references to the Arbitrage Pricing Theory (APT), Black-Scholes-Merton model (BSM), Capital Asset Pricing Model (CAPM) and Modern Portfolio Theory (MPT). We refer to the list of monographs in the references for notations and suitable background information.

The Sharpe ratio appears here very frequently, so let us recall it for convenience:
\begin{equation}\label{eq: sharpe}
\mathrm{Sharpe\ ratio} = \frac{\E(R-R_f)}{\sqrt{\mathrm{Var}(R)}}
\end{equation}
where $R$ is the stochastic \emph{rate} of return of an asset in question and $R_f$ is the rate of return of a benchmark risk-free asset, both annual. 

Let us recall some well-known formulas relevant to the discussion for the sake of convenience. 
Let us calculate the ratio discussed in the introduction. First, let us recall the density functions 
of the physical measure $\P$ and risk neutral measure $\Q$ in the BSM model. Here we are particularly 
interested in the relative increments $\frac{S_t}{S_0}$ in the value of the equity where 
$S_0$ is deterministic. The $\P$-density is given by 

\[\P(S_t/S_0 \in A)=\frac{1}{\sigma\sqrt{2\pi t}}\int_{A}\frac{1}{y}e^{-\frac{(\ln(y/S_0) -(\mu-\sigma^2 /2)t)^2}
{2\sigma^2 t}}\ dy,\]

that is, $\ln(S_t/S_0)$ considered with respect to $\P$ measure has the law\\ 
\mbox{$N((\mu- \frac{1}{2}\sigma^2) t,\ \sigma^2 t)$}. 

The risk neutral density is similar, only the center of the distribution is shifted downwards:

\[\Q(S_t/S_0 \in A)=\frac{1}{\sigma\sqrt{2\pi t}}\int_{A}\frac{1}{y}e^{-\frac{(\ln(y/S_0) -(r-\sigma^2 /2)t)^2} {2\sigma^2 t}}\ dy\]
(see e.g. \cite[p. 269]{Follmer}), that is, $\ln(S_t/S_0)$ considered with respect to $\Q$ measure has the law
\mbox{$N((r- \frac{1}{2}\sigma^2) t,\ \sigma^2 t)$}.

The price of a European call at time $t$ having payoff $(S_T - K)^+ :=\max(S_T - K,0)$ at maturity $T$ 
is denoted by $C(S_t, t,T,K)$. The following equation holds:
\begin{equation}\label{eq: call}
\begin{split}
\E_{\Q}(S_T - cS_t)^+ &=\frac{1}{e^{r(T-t)}\sigma\sqrt{2\pi (T-t)}}\int_{cS_t}^{\infty}\frac{y-cS_t}{y}e^{-\frac{(\ln(y/S_t) -(r-\sigma^2 /2)(T-t))^2} {2\sigma^2 (T-t)}}\ dy
\\
&=\frac{S_t}{e^{r(T-t)}\sigma\sqrt{2\pi (T-t)}}\int_{c}^{\infty}\frac{x-c}{x}e^{-\frac{(\ln(x) -(r-\sigma^2 /2)(T-t))^2} {2\sigma^2 (T-t)}}\ dx\\
&=C(S_t,t,T,cS_t)
\end{split}
\end{equation}
for all constants $c>0$ by change of variable $x=y/S_t$. Similarly we obtain that 
\begin{equation}\label{eq: call2}
\E_{\P}(S_T - cS_t)^+ =\frac{S_t}{\sigma\sqrt{2\pi (T-t)}}\int_{c}^{\infty}\frac{x-c}{x}e^{-\frac{(\ln(x) -(\mu -\sigma^2 /2)(T-t))^2} {2\sigma^2 (T-t)}}\ dx.
\end{equation}
Denote the above expectation by $S_t E_c$. Then we may write the $\P$-standard deviation of the call payoff
as follows:
\begin{multline*}
\sqrt{\E_{\P}((S_T - cS_t)^+ - S_t E_c)^2}\\
=\left(\frac{1}{e^{r(T-t)}\sigma\sqrt{2\pi (T-t)}}\int_{0}^{\infty} \frac{((y-cS_t)^+ - S_t E_c)^2}{y}
e^{-\frac{(\ln(y/S_t) -(\mu-\sigma^2 /2)(T-t))^2} {2\sigma^2 (T-t)}}\ dy\right)^{\frac{1}{2}}\\
=S_t \left(\frac{1}{e^{r(T-t)}\sigma\sqrt{2\pi (T-t)}}\int_{0}^{\infty}\frac{((x-c)^+ - E_c)^2}{x}
e^{-\frac{(\ln(x) -(\mu-\sigma^2 /2)(T-t))^2} {2\sigma^2 (T-t)}}\ dx\right)^{\frac{1}{2}}.
\end{multline*}

\section{The mechanism described}
\subsection{The ratio}
Recall that the state price density and stochastic deflator can be written by means of the Radon-Nikodym derivative which can be computed as follows:
\begin{equation}\label{eq: ratio1} 
\frac{d\Q}{d\P}\bigg|_{\ln (S_\tau / S_0)=x}=\frac{e^{-\frac{(x -(r-\sigma^2 /2)\tau)^2}{2\sigma^2 \tau}}}{e^{-\frac{(x -(\mu-\sigma^2 /2)\tau)^2}{2\sigma^2 \tau}}}
\end{equation}
and straight forward algebraic manipulations yield that \eqref{eq: ratio1} equals
\begin{equation}\label{eq: ratio2}
=e^{(r-\mu)(\frac{x}{\sigma^2}-\tau(r+\mu)/2\sigma^2 + \tau/2)} .
\end{equation}
Recall that $r-\mu<0$ (typically). This can be seen rather a structural property than empirical fact, since in reasonable models the expected rate of returns of risky assets are higher than that of the risk-free assets.

We see immediately that \eqref{eq: ratio2} tends to $0$ as $x\to \infty$ 
and tends to $\infty$ as $x\to -\infty$. Since $(r+\mu)/\sigma^2 >1$ (typically), we have that
\begin{equation}\label{eq: ratio_lim_3}
e^{(r-\mu)(\frac{x}{\sigma^2}-\tau(r+\mu)/2\sigma^2 + \tau/2)} \searrow e^{\frac{x(r-\mu)}{\sigma^2}}
\end{equation}
as $\tau\to 0$. This quantity has the same asymptotics with respect to $x$ as \eqref{eq: ratio2}
and decays exponentially as $x$ grows.
By definition, the right hand side of \eqref{eq: ratio_lim_3} equals to $1$ when $\ln(S_\tau / S_0)=x=0$, 
that is, the critical point where physical probability and the risk-neutral one asymptotically coincide, is at-the-money. 
The role of $\sigma$ in this context will be discussed subsequently. In that connection we will apply the short time scale ratio $e^{\frac{x(r-\mu)}{\sigma^2}}$, since it is convenient to work with.

\subsection{The trading strategy}

An essential feature of the trading strategy discussed shortly is that we would like to buy very cheap, out-of-the-money European style 'plain vanilla' calls with short horizon. Then we wait to the end of the horizon, and, regardless of the outcome, we buy the next patch, wait, and keep doing this, 
say, to the end of the year. In practice the lengths of intervals between expiration dates of traded European style options are bounded from below, 
for example for index options on S\&P 500 it is one month. We are also required to trade very small amounts of calls. 

However, in the BSM model there is no structural constraint preventing us from buying and selling options every second
worth a cent. We will follow the BSM framework in other aspects too. Thus, we will assume that there are no 
transaction costs, liquidity concerns, or any other such market imperfections. We will assume that when cash 
is not invested in calls, it is invested in risk-free bonds recognized by the BSM model. We will assume that the parameters of the BSM model remain constant during the whole time. 
In the next section we will analyze a yet simpler situation for the sake of transparency of the ideas present.

The downside of the theoretical considerations proposed here, of course, is that were are dealing with phenomena appearing only in micro scale when we push the BSM model to the limit. This happens in the model when buying extremely deep out-of-the-money calls. 

Although the assumptions made are very strong with potential practical applications in mind and the situations 
arising can be considered as marginal, we suspect that the fact that 'the odds are not against' the proposed strategy could potentially pave way for some suitable automated options trader detecting bargains. 
This is to say, we believe the idea presented here could be combined with some other high-frequency trading functions. The practical problem with finding extremely deep out-of-the-money options and the limited number of expiries could be somewhat relaxed by internationally diversifying to all the European type calls. In fact, as it turns out in the next section, the high Sharpe ratio resulting from running the investment strategy is not restricted to using plain vanilla calls.

Next we will describe our trading strategy. In order to make sure it is self-financing, we will
first fix the amount to be invested, say $I$ units of numeraire during $1$ year. We will divide the year 
equally to $n$ time intervals $[t_{i},t_{i+1}],\ i\in \{0,1,\ldots,n-1\},\ t_0 =0,\ t_n =1$. 
Since the model parameters $\mu,\sigma$ and $r$ are constant over time, an inspection of the option price formula 
\eqref{eq: call} reveals that there exists a unique constant $c=c_{n,I}>0$ such that 
\begin{equation}\label{eq: c}
C(S_{t_i},t_{i},t_{i+1},cS_{t_i})=I S_{t_i}/n\quad \mathrm{for\ all}\ i\in \{0,1,\ldots,n-1\},
\end{equation}
regardless of what the realized values $S_{t_i}$ shall be. 
This is a crucial fact and it makes the strategy simple to implement and to analyze. 

The algorithm of the investment strategy is as follows:
\begin{enumerate}
\item[(i)]{At time $t=0$ we have $I$ units of numeraire is at our disposal.}
\item[(ii)]{At time $t_0$ we buy $1/S_{t_0}$-many European calls maturing at time $t_{1}$ with strike $c S_{0}$. Here $c$ is the constant appearing in \eqref{eq: c}.}
\item[(iii)]{At time $t_1$ we invest the possible proceeds of the maturing call to risk-free bonds.}
\item[(iv)]{At time $t_1$ we buy $1/S_{t_1}$-many European calls maturing at time $t_{2}$ with strike $cS_{t_1}$.}
\item[(v)]{We continue in this manner to the end of the year $t_{n}$, at which point we have the yield of 
the tradings invested in risk-free bonds.}
\end{enumerate} 

Because of the $1$-to-$1$ correspondence (for a fixed $n$) of $I$ and $c_{n,I}$ we may alternatively 
describe the strategy by \emph{first} choosing  the constant $c$ and then solve the initial capital $I$ to be invested accordingly.

Since we buy a portfolio of $1/S_{t_i}$-many calls with strike $cS_{t_i}$ at time $t_i$ we observe similarly as in \eqref{eq: call} and \eqref{eq: call2} that the yield of such a call portfolio has essentially the same risk-neutral and physical distributions as the yield of one call with parameters $K=cS_i$, $S_i=1$ and the same running time interval. 
Thus, since the increments $S_{t_{i+1}}-S_{t_i}$ are independent in the BSM model, we are only required to analyze the expected value, variance and the price $C(S_t,t,T,K)$ of one call only, it is convenient by virtue of
\eqref{eq: call} and the preceding formulas to assume that 
\begin{equation}\label{eq: st1}
S_t=1.
\end{equation}

\subsection{Numerical illustrations}

Some relevant values of the cash flows resulting from running the strategy are presented below. For the sake of simplicity we will begin by 
choosing different values for the constant $c$, instead of choosing  the initial wealth level in the investment strategy. 
The computations are done by using Matlab with the BSM model parameters $\mu=0.1$ (trend), $r=0.04$ (short rate), 
$\sigma=0.20,\ 0.30,\ 0.40,\ 0.50$ (implied volatility) and strike prices $K=cS_{t}$ with ($S_t =1$ nominally by
\eqref{eq: st1}) $c=1+0.005\times j$, $j=1,\ldots,50$ (number of steps above at-the-money) and $n=12$ (months). The following figures illustrate the fact that the trading strategy is most lucrative (on paper) and is sensitive to the availability of deep out-of-the-money calls and large number of holding periods. 


\begin{figure}[H]
\centering
\includegraphics[width=10cm]{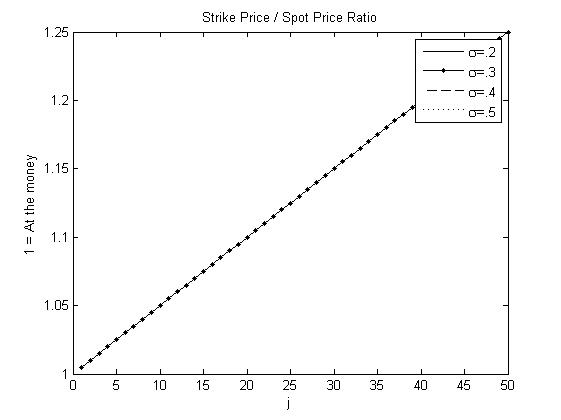}
\caption{The value of constants, $c=1+0.005\times j$, appearing in the trading algorithm in different scenarios.}
\end{figure}

\begin{figure}[H]
\centering
\includegraphics[width=10cm]{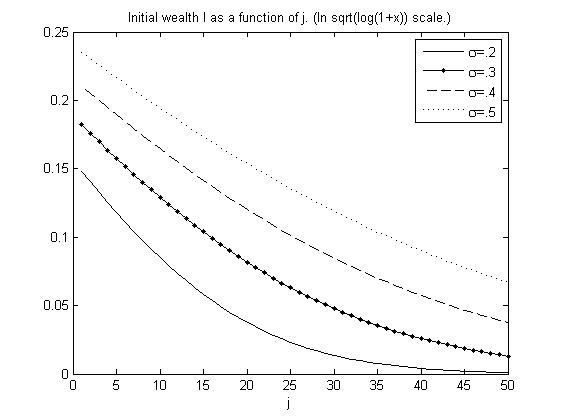}
\caption{Initial wealth in different cases; implied volatilities $\sigma= 0.2 , \ldots, 0.5$.
Scale motivated by the function $e^{-x^2}$ in the Gaussian density.}
\end{figure}

\begin{figure}[H]
\centering
\includegraphics[width=10cm]{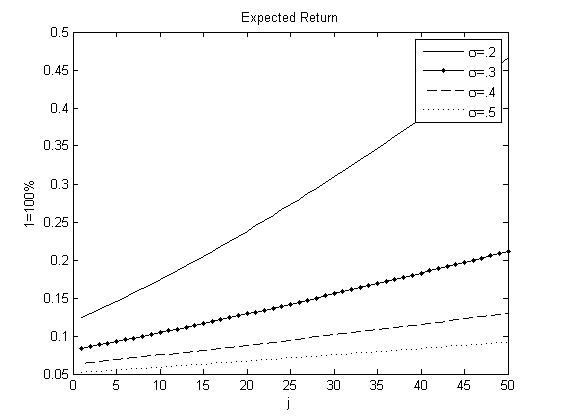}
\caption{The expected returns corresponding to different volatilities and strike price steps $j$. Here $0\%$ corresponds to braking even on average.}
\end{figure}

\begin{figure}[H]
\centering
\includegraphics[width=10cm]{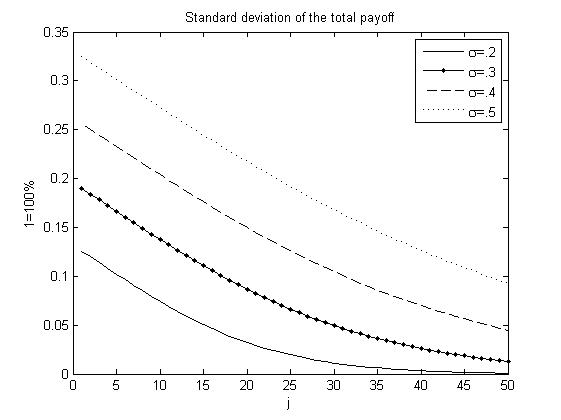}
\caption{}
\end{figure}

\begin{figure}[H]
\centering
\includegraphics[width=10cm]{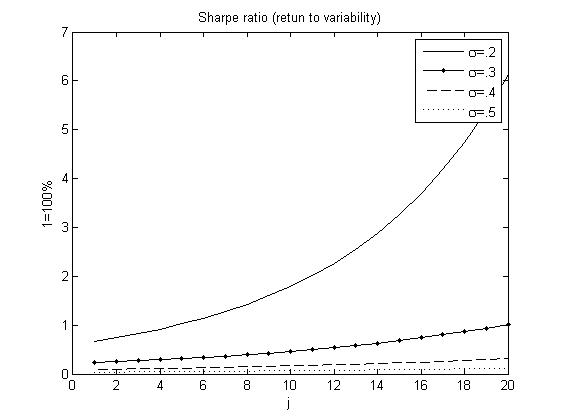}
\caption{The Sharpe ratio of the annual cash flow resulting from the trading strategy. For comparison, a good Sharpe ratio for a large base equity portfolio is 
of the magnitude $1$.}
\end{figure}

\begin{figure}[H]
\centering
\includegraphics[width=10cm]{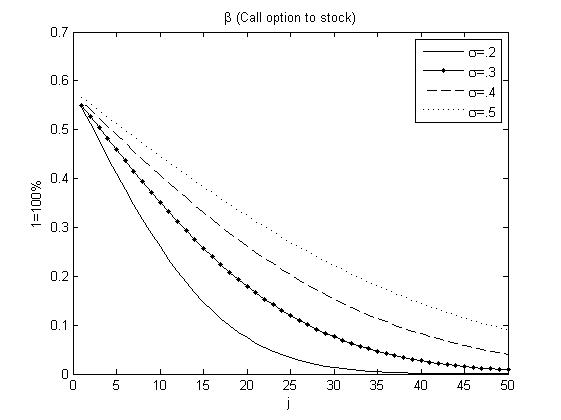}
\caption{}
\end{figure}

These computations suggest that if one is to expect BSM type conditions on the market for a year
(e.g. constant volatility, no autocorrelation on the equity returns, normally distributed increments and no volatility smile),
then buying a cheap call every month already results in high Sharpe ratios and low beta.

\begin{figure}[H]
\centering
\includegraphics[width=10cm]{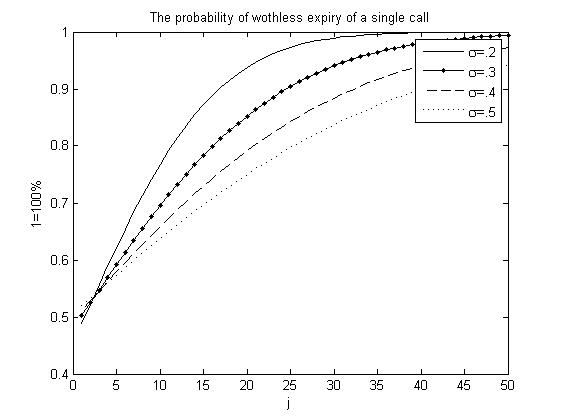}
\caption{}
\end{figure}

\begin{figure}[H]
\centering
\includegraphics[width=10cm]{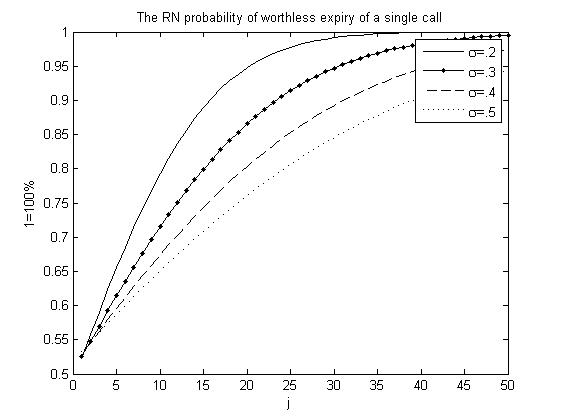}
\caption{}
\end{figure}

\begin{figure}[H]
\centering
\includegraphics[width=10cm]{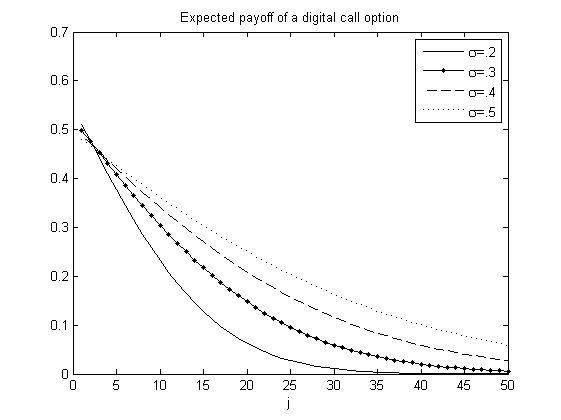}
\caption{}
\end{figure}

\begin{figure}[H]
\centering
\includegraphics[width=10cm]{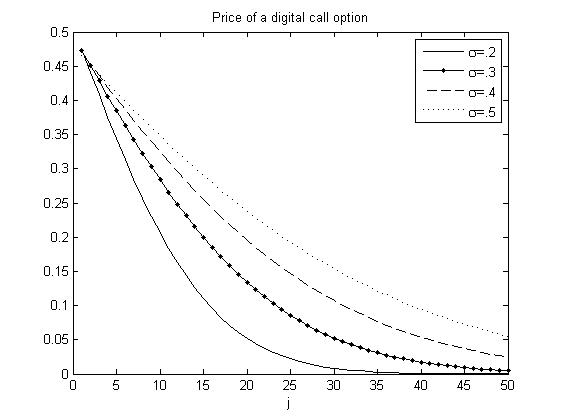}
\caption{}
\end{figure}

\begin{figure}[H]
\centering
\includegraphics[width=10cm]{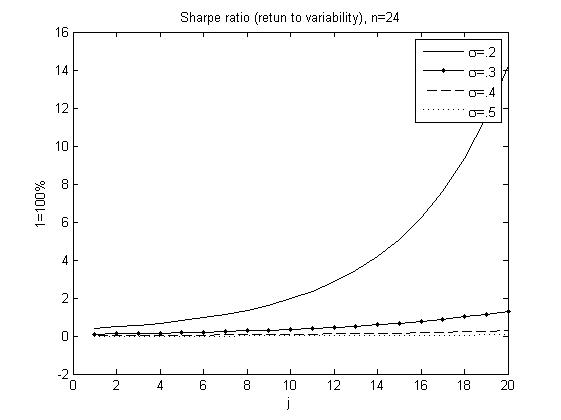}
\caption{Increasing the number of holding periods p.a. improves the Sharpe ratio.}
\end{figure}

\section{Theoretical considerations}
\subsection{Digital options}

We will take a look at digital options since they are easy to analyze and yet they capture the essential phenomenon 
discussed here. Since the analysis is transparent and even rather elementary with these options, the 
upside is that we do not require any numerical computations. Thus we will revisit the trading strategy by using 
digital options in place of the plain calls.

Let us consider the price $D(S_t, t, T, K)$ of a European style digital option, which pays $1$ unit of numeraire 
at the time of maturity $T$ if the underlying security satisfies $S_T \geq K$. This kind of digital option can be 
approximated, both in payoff and price, by buying $i$ many European options with strike price $K$ and shorting $i$ many 
options with strike price $K+1/i$, see \cite{BL} and \cite{Jarrow}. Here the time to maturity for all the options mentioned is the same and the approximation improves as $i$ increases.\footnote{It is useful to think of the definition of differential and put $h=1/i$, $i$ being a large natural number.}  

In the BSM model the value of the digital option at time $t$ is 
\[D(S_t, t, T, K)=e^{-r(T-t)}\E_{\Q}(1_{S_T \geq K})=e^{-r(T-t)} \Q(S_T \geq K)\]
where the current value of $S_t$ is deterministic. 
The risk-neutral probability on the right hand side equals 
\[\Q(S_T \geq K)=\frac{1}{\sigma\sqrt{2\pi (T-t)}}\int_{K}^{\infty}\frac{1}{y}e^{-\frac{(\ln(y/S_t) -(r-\sigma^2 /2)(T-t))^2}{2\sigma^2 (T-t)}}\ dy.\]
In contrast, this should be compared to the physical probability that the option is triggered:
\[\frac{1}{\sigma\sqrt{2\pi (T-t)}}\int_{K}^{\infty}\frac{1}{y}e^{-\frac{(\ln(y/S_t) -(\mu-\sigma^2 /2)(T-t))^2}{2\sigma^2 (T-t)}}\ dy.\]

In the investment strategy of the previous section we kept cash invested in risk-free bonds when 
not invested in calls. However, the accumulation of interest does not play a significant role 
in this paper. Therefore, in what follows we will instead consider cash being invested in a bank account with zero interest
for the sake of simplicity, and, similarly, we will treat the discount factor $e^{-r(T-t)}$ as $1$ by convention. 
Indeed, this causes no problem as we are not running any kind of replication strategy to synthesize derivatives and therefore we may separate the short rates appearing in the BSM model and in the bank account. The Sharpe ratio \eqref{eq: sharpe} is invariant under positive scaling of cash flows, therefore in calculations it does not matter what is the initial capital to be invested.
  
The expected rate of return for a digital option is easy to formulate: 
\begin{equation}\label{eq: int_ratio}
R=\frac{\int_{K}^{\infty}\frac{1}{y}e^{-\frac{(\ln(y/S_0) -(\mu-\sigma^2 /2)(T-t))^2}
{2\sigma^2 (T-t)}}\ dy}{\int_{K}^{\infty}\frac{1}{y}e^{-\frac{(\ln(y/S_0) -(r-\sigma^2 /2)(T-t))^2}{2\sigma^2 (T-t)}}\ dy} -1.
\end{equation}
The above ratio reads as follows: the potential payment of the option (i.e. $1$) times the probability 
of the option being triggered divided by the purchase price of the option. 
Since \eqref{eq: ratio2} is decreasing in $x$, it is easy to check that the fraction in 
\eqref{eq: int_ratio} satisfies
\begin{equation}\label{eq: R_geq}
R =\frac{p}{q}-1 \geq e^{(\mu-r)(\frac{\ln(K/S_t)}{\sigma^2}-(\mu+r)/n2\sigma^2 + 1/2n)} -1
\end{equation}
where $1/n$ is the length of each holding period. In particular, the expected return on investment 
\eqref{eq: int_ratio} tends to infinity as $K$ increases.

Now, suppose that we will use digital options in a similar fashion as in the previously introduced trading strategy. 
Then the outcome is easy to understand. Namely, for $n$ holding periods $i=0,1,\ldots ,n-1$ we will buy 
in the beginning of each period $1/S_{t_i}$ digital options with horizon $1/n$ and strike $c_i S_{t_i}$ such that the 
total worth is $1/n$. Then the payoff is binomially distributed, since we are essentially dealing with an i.i.d. sequence of binary random variables. The expected payoff of each of the independent bets is given by
\begin{equation*}
\begin{split}
p_i=\P(S_{t_{i+1}}\geq c_i S_{t_i})&=\frac{1}{\sigma\sqrt{2\pi /n}}\int_{c_i S_{t_i}}^{\infty}\frac{1}{y}e^{-\frac{(\ln(y/S_{t_{i}}) -(\mu-\sigma^2 /2)/n)^2}{2\sigma^2 /n}}\ dy\\
&= \frac{S_{t_i}}{\sigma\sqrt{2\pi /n}}\int_{c_i}^{\infty}\frac{1}{x}e^{-\frac{(\ln(x) -(\mu-\sigma^2 /2)/n)^2}{2\sigma^2 /n}}\ dx.
\end{split}
\end{equation*}
where $c_i$ is the unique solution to the equation
\[1/n=\frac{q_i}{S_{t_i}}=\frac{1}{\sigma\sqrt{2\pi /n}}\int_{c_i}^{\infty}\frac{1}{x}e^{-\frac{(\ln(x)  -(r-\sigma^2 /2)/n)^2}{2\sigma^2 /n}}\ dx.\]
Above we applied the change of variable $x=y/S_{t_i}$. We observe that $\frac{p_i}{q_i}\to \infty$ as $c_i\to \infty$. In the appendix we will show 
that the latter holds as $n\to \infty$.

The expected payoff resulting from running the above strategy with binary options is clearly $\sum_i  p_i$ and the standard deviation
of the payoff is $\sqrt{\sum_i p_i (1-p_i)}$. The Sharpe ratio is then
\begin{equation*}
\begin{split}
&\frac{\sum_i p_i -1}{\sqrt{\sum_i p_i (1-p_i)}}\geq \frac{\sum_i p_i -1}{\sqrt{\sum_i p_i}} \to \sqrt{\sum_i p_i }\\
=&\sqrt{\frac{1}{n}\sum_i \frac{p_i}{q_i}}\to \infty,\quad n\to \infty.
\end{split}
\end{equation*}

We note that choosing high $c$ and $n$ results in low $p$ and $q$, and high $K$, $R$ in \eqref{eq: R_geq}. 
In the above trading strategy we let the parameter $c$ (and thus $K$) vary in order to circumvent some technical calculations involving \eqref{eq: R_geq}. 

We note that it is essential above that we let $K$ and $p/q$ tend to infinity. 
Namely, for a fixed $p/q$ the corresponding  binomial payoff process Sharpe ratio tends to $0$.

\subsection{Double digital options}
The above considerations also apply if one uses double digital options, instead of simple digital ones. 
Recall that a double digital option is a European style derivative which pays of $1$ unit of numeraire if the 
value of the underlying asset hits a closed interval at maturity.

The above strategy implemented with double digital options also results in binomially distributed cash flow. 
For short intervals, where the option is triggered, the ratio $q/p$ will be close to
\eqref{eq: ratio2}, hence easy to understand. 

For infinitesimal intervals the double digital options can be viewed as theoretical 
Arrow-Debreu securities and in this case the ratio \eqref{eq: ratio2} holds exactly for $q/p$.

\subsection{Incompatibility of the BSM model and the CAPM}
The existence of the described (theoretical) investment opportunities appears not to align well with 
the Modern Portfolio Theory, since the Sharpe ratio of the investment opportunity grows unrealistically high. 

The presented trading strategy has also some convenient linear statistical properties following from the 
simple feature that \emph{most of the time} the European calls bought expire \emph{worthless}. Namely, because of 
this reason and the basic properties of linear correlation the yield of the strategy has beta close to zero for large $n$.
Roughly speaking, in most holding periods the payoff is completely uncorrelated with the market.
This remains true even if the underlying asset was a large base equity index corresponding to the market portfolio. 
The picture is, however, complicated by the fact that the investment strategy payoff process moves with large increases of the underlying asset.
See appendix for information about the application of the central limit theorem to analyze the beta.

In the BSM framework the successively held options are non-autocorrelated and even independent. According to \eqref{eq: call2} 
and the fact that we buy $1/S_{t}$-many calls, the payoffs resulting from each of the holding periods are identically distributed.
Therefore the total proceeds of the investment strategy are roughly normally distributed by the Central Limit Theorem 
for a large number of holding periods\footnote{The obvious distortions in distributions resulting e.g. from the limited losses are shared by models of equity value. } 

Since the beta can be made very small, the theoretical investment opportunity is a fortiori incompatible with CAPM. 
The BSM model suggests singularly lucrative investment opportunities which cannot be fitted in the CAPM framework, 
i.e. simultaneously low beta and high Sharpe ratio. Thus the models appear structurally incompatible. Essentially same considerations apply to the APT model, 
since it is difficult to predict \emph{very specific} short-term fluctuations of the stock prices or equity index by macroeconomic factors.

It is not clear how this mismatch should be interpreted. Perhaps these findings suggest that the 
equilibrium valuation models (e.g. CAPM and APT) prone to cancellations of risk, and, on the other side, 
the hedging-based valuation, represent different paradigms of financial economics. 
For example, the term 'arbitrage' is a very different notion in the context of risk-neutral pricing, 
compared to its occurrence in the APT where the term rather stands for statistical arbitrage.
In the risk-neutral pricing one identifies the values of cash flows which coincide with probability $1$. On the other hand, 
in APT one identifies values of cash flows having only the same distribution, in fact only certain same parameters (factors) describing the distribution.

\subsection{Volatility smile}
As we already pointed out in the introduction, the investment strategies are based on the 
key observation that the physical and risk-neutral densities are not asymptotically comparable. 
We have so far discussed strategies in which one buys successively European vanilla or digital \emph{calls} 
but the non-comparability of the densities at the lower tail suggests an analogous investment strategy where one 
successively shorts puts instead.

Suppose that the proposed investment strategy type were viable in practice. 
Then a wide adoption of these trading strategies should adjust upward the market prices of calls out-of-the-money
and with a short horizon. Similarly, the market prices of short-horizon puts with low strikes should adjust downward, compared to the benchmark provided by the BSM model.\footnote{Intuitively, this appears to result 
in lower implied volatility but the following formal reasoning suggests the opposite.} If the volume of the underlying security is outnumbered by the quantity of liquid options traded, then our attention in pricing the options shifts from hedging arguments to the law of supply and demand, cf. [Fengler 2005, p.45].

Next we will see what happens when the state price density curve is adapted into a shape such that the 
proposed investment strategy ceases to exist. Let us analyze the asymptotic ratio \eqref{eq: ratio_lim_3}, since 
it involves a simple relationship between the strike price and the volatility. Namely, the situation where
\[\frac{d\P}{d\Q}\approx e^{\frac{x(\mu -r)}{\sigma^2}}\]
tends to $\infty$ (respectively $0$), as $x\to \infty$ (respectively $x\to -\infty$) results in option prices incompatible
with some equilibrium valuation models, like CAPM. One consistent remedy to this situation 
is to change the underlying state price density (see also \cite{Fengler}). 
Let us replace the constant $\sigma$ in \eqref{eq: ratio_lim_3} by a function $\sigma(x)$. 
Our heuristic rationale is that $\sigma(x)$ should be chosen in such a way that
\begin{equation}\label{eq: R} 
R_{-}< \frac{x(\mu-r) }{\sigma(x)^2}< R_{+}
\end{equation}
for some reasonable (finite) bounds $R_{-} < 0< R_{+}$. This means that the BSM state price density could by
adjusted by varying $\sigma$ continuously in such a way that at-the-money no change occurs 
(see remarks after \eqref{eq: ratio_lim_3}) and in the above inequality the bounds are obtained asymptotically.

We leave the exploration of these bounds for future research but we suggest that studying the Sharpe ratios of the resulting investment strategies' payoffs should be a good starting point. 
 
The above inequalities suggest the following asymptotics for $\sigma(\ln(K))$ in adjusting the state price density, 
which also involves the volatility smile:
\begin{equation}
\sigma(\ln(K)) \approx \left\{ \begin{array}{rl}
 \sqrt{\frac{\ln(K)(\mu-r)}{R_{+}}} &\mbox{ for $0< < \ln K$} \\
 \sqrt{\frac{\ln(K)(\mu-r)}{R_{-}}} &\mbox{ for $\ln K < < 0$}
       \end{array} \right.
\end{equation}

Next we will give as an example an \emph{ad hoc} formula for volatility smile curve.
We choose $R=R_{+}=-R_{-}$ and apply a suitable logistic-like function to model the above asymptotics.
Put 
\[L(x)=\sign(x)(1-e^{-\sqrt{\abs{x}}})R\]
and consider
\[\frac{x(\mu-r)}{\sigma_1 (x)^2}=L(x)\]
where $\sigma_1 (0)=0$. We obtain the following heuristic volatility smile formula for $x=\ln(K /S_0)$:
\begin{equation}\label{eq: smile}
\begin{split}
\sigma(x)&=\sigma_0 + \sqrt{\frac{x(\mu-r)}{L(x)}}, \quad x\neq 0,\\
\sigma(0)&=\sigma_0 .
\end{split}
\end{equation}
This approach, although being ad hoc and not producing a completely satisfying volatility smile curve, 
is rather transparent in what comes to describing the tails. 

We have discussed a trading strategy with short time intervals, which then amplifies 
the phenomenon under investigation. This suggests that the above explanation of the volatility 
smile should be most relevant for options having short time to maturity and low base volatility. 
The constant $R$ above appears like a market smile factor which should be calibrated from the market data. 

In fact, the factor $R$ has a clear interpretation in our setting. Namely, since $R$ was 
chosen to be the supremum of the absolute value of the inverse of the right hand side of \eqref{eq: ratio_lim_3},
the quantity $\exp(R)-1$ can be described as the least upper bound for the expected rate of return for all 
European style derivatives in the model with short horizon. In the above framework the feature that markets 'tolerate' high $R$ corresponds to weak volatility smile.

\section{Discussion}

We have used structural assumptions of the BSM model to argue that some form of volatility smile 
is an expected phenomenon. Surely we are not claiming here that fat tails of return distributions and jumps
do not contribute to volatility smile.

We make next some remarks. 
The latter example with digital options produces expected returns growing slower because in treating regular calls
the integration of the function $(x-K)^{+}$ (instead of $1_{x\geq K}$) against both physical and risk-neutral 
probabilities puts more weight on the upper tail where the ratio $\frac{d\P}{d\Q}(\ln K)$ increases rapidly. 
Thus the discussed phenomenon amplifies in the case with the European plain vanilla calls (instead of digital options).

We argued the beta of the payoff of the investment strategy is small since most of the time the calls bought expire worthless. This intuitively means that in most holding periods the payoff happens to be completely uncorrelated with the market. Similarly, the autocorrelation of the payoff process is close to zero, even if the underlying were strongly autocorrelated. Also, according to the same reasoning the correlation of the yield process with macro economic indicators employed by the APT model is close to zero.

\subsection{Extensions}  
The basic phenomenon discussed does not change considerably if the parameters $\mu$, $r$ and $\sigma$ change over time. 
Therefore the strategy should run similarly with similar conclusions even if the model is updated with new parameter values 
in between the holding periods.

There are of course issues not considered here related to deviation form the BSM framework, such as the non-normality 
of the yield distribution. Perhaps the issue with the distributions is not crucial here because we essentially applied 
only the very rudimentary property of the pricing system, namely that $\frac{d\Q}{d\P}|_{S_T = K}$  tends to $0$ as $K\to \infty$. 
Therefore the strategy should be successfully executable in any pricing systems with this key feature. 
Immediately two questions arise next:
\begin{enumerate}
\item{What happens if one uses other than European style of options in the investment strategy?}
\item{What happens if one trades European calls with similar logic but does not wait to maturity?}
\end{enumerate}

These problems are related to the fact that the times to expiry of issued options are bounded from below
so that in practice one cannot take $n$ to be very large annually.

There are two more serious issues in the real markets that appear to impair the functioning of the described strategy.
One is market information implicit in the prices of derivatives and the other is autocorrelation of the 
securities. 

Often the prices of derivatives are considered to contain information (or at least market sentiment) 
about the future values of the underlying. Therefore one could argue that the described strategy fails because
the markets anticipate the future rise/drop of the security value in the derivative price 
and thus the effect of true randomness strongly exploited here vanishes. Moreover, one 
could argue that even though in the long run the future values of the market are hard to predict, the market 
has disproportionately good insight about the near future. This could appear to be a plausible worry, 
especially in the sense that we would possibly end up betting against well-informed agents. 
Also, the markets could react to such consistent trading by making deep-out-of-the-money calls less cheap.
We do not only restrict our concern to a `spontaneous' response of the markets due to the law of supply and demand but possibly by a strategic decision of some agents as well. This was already discussed in connection to volatility smile.  
 
Another possible critique involves autocorrelated securities returns. Namely, keeping betting 
(up, in our setting) in a bearish trend could have similar consequences as betting against well-informed investors.
We admit that the presented strategy is probably not such a good idea in a bearish market.

Still, there appears to be a possible remedy to these suggested problems. In both the 
examples the problem was a severe failure of the models resulting from a kind of lack of of randomness. 
The idea is to cultivate the strategy further in order to forcibly randomize events, in a sense. 

Let us first try to deal with the informed markets effect. We claim that even if some agents participating
in the markets have information about the future values of the underlying security or index, it is 
not easy to predict the \emph{exact} value of the underlying. Therefore we can circumvent the proposed problem 
by trading double digital options instead of plain vanilla calls. The payoff function of a double digital option 
can be regarded as an indicator function $1_{[a,b]}$ on possible short interval $[a,b]$. The shorter the interval, 
the more difficult it is obviously for anyone, no matter how informed, to predict if the underlying is going 
to hit the interval at the date of maturity or not. The interval can be placed far in the upper tail, so that the expected returns tend high according to \eqref{eq: ratio2}.

In what comes to the problem with the autocorrelation, one possible, rather theoretical approach is to attempt 
a similar randomization method by making the step sizes ultra-short and then executing trades only in every tenth interval, 
or so.  

We already discussed how the trading strategy works with digital options. These can be regarded as power options 
of power $0$ under the conventions $(\max(S_T - K,0))^{0}=1$ if $S_T>K$ and $0^{0}=0$. 
Since the limit \eqref{eq: ratio_lim_3} decays exponentially, we could instead use power options $(\max(S_T - K,0))^{p}$ 
of any power $0\leq p <\infty$ in the trading strategy. Note that for $p>1$ and large values of $K$ 
the upper tail values have relatively higher weight than in the case with regular calls and therefore 
the effect on expected returns becomes more pronounced.

Expected returns are more sensitive to the choice of the constant $c$, compared to the number of holding periods $n$. For low levels of implied volatility $\sigma$
one can choose the constant $c$ closer to $1$ compared to the case with higher volatilities.
In \eqref{eq: ratio_lim_3} we considered the case where 
\[\frac{\mu + r}{\sigma^2}>1.\]
It could easily happen that the above quotient is $\leq 1$ as well, for example in our case with the constants $\mu=0.1$, $r=0.04$ and $\sigma\geq 0.37...$.
The critical volatility $\approx 0.37$ can be seen as an inversion level where the effect of the time span $\tau$ on the ratio \eqref{eq: ratio_lim_3} changes its direction ($\pm$).
Interestingly, for this critical volatility the short time scale ratio $e^{\frac{x(r-\mu)}{\sigma^2}}$ is the correct value for \eqref{eq: ratio2}, i.e. the ratio remains the same 
in  all time scales. 

Let us consider the extreme cases in \eqref{eq: ratio_lim_3} involving the value $\frac{\mu + r}{\sigma^2}$ far away from the inversions level, that is, low implied volatility and high implied volatility, e.g. $\sigma=0.20,\ 0.50$, respectively. The above observation about the change of the effect of the time scale suggests the following, in principle testable hypothesis involving volatility smiles. Namely, for shortening the expiry of out-of-the-money calls increases the expected return of the call in the case of low overall implied volatility, 
ceteris paribus, whereas it decreases the expected return of the call in case of high overall implied volatility.

\subsection*{Appendix}

\subsubsection*{The strike to spot ratio $c$ of the trading strategy}

Recall the formula
\[1/n=\frac{q_i}{S_{t_i}}=\frac{1}{\sigma\sqrt{2\pi /n}}\int_{c_i}^{\infty}\frac{1}{x}e^{-\frac{(\ln(x)  -(r-\sigma^2 /2)/n)^2}{2\sigma^2 /n}}\ dx.\]

Write 
\[1/s = \frac{1}{\sigma\sqrt{2\pi /s}}\int_{c(s)}^{\infty}\frac{1}{x}e^{-\frac{(\ln(x)  -(r-\sigma^2 /2)/n)^2}{2\sigma^2 /n}}\ dx,\quad s > 1 .\]

We wish to check that $c(s)\to \infty$ as $s\to \infty$.

Note that
\[\frac{1}{\sqrt{s}} = a_1  \int_{c(s)}^{\infty}\frac{1}{x}e^{-\frac{(\ln(x)  -(r-\sigma^2 /2)/n)^2}{2\sigma^2 /n}}\ dx.\]
Observe that $c(s)$ must be a unique function which is also strictly increasing and continuous.

Assume first that $c(s)$ is continuously differentiable. Then 
\[-\frac{1}{2} s^{-\frac{3}{2}} = - a_1 \frac{c' (s)}{c(s)}e^{-\frac{(\ln(c(s))  -(r-\sigma^2 /2)/n)^2}{2\sigma^2 /n}}.\]
Actually, from this form one can deduce easily a continuously differentiable solution $c(s)$, thus it is continuously differentiable by uniqueness.
From the above form we observe that $c'(s)>0$ for all $s>1$. We obtain 
\begin{equation*}
\begin{split}
1 &= a_2  c' (s) e^{-\frac{(\ln(c(s))  -(r-\sigma^2 /2)/n)^2}{2\sigma^2 /n}+\frac{3}{2}\ln s - \ln c(s)}\\
&\approx a_2 c'(s) e^{-s(\ln c(s))^2 /2\sigma^2 - \ln c(s) (r-\sigma^2 /2)/\sigma^2 +\frac{3}{2}\ln s - \ln c(s)}
\end{split}
\end{equation*}
for large $s=n$. If $c(s)$ approaches a real number asymptotically, then 
\[-s(\ln c(s))^2 /2\sigma^2 - \ln c(s) (r+\sigma^2 /2)/\sigma^2 +\frac{3}{2}\ln s \to - \infty,\quad s\to \infty,\]
thus 
\[e^{-s(\ln c(s))^2 /2\sigma^2 - \ln c(s) (r+\sigma^2 /2)/\sigma^2 +\frac{3}{2}\ln s }\to 0,\quad s\to \infty\]
which means that 
$c'(s)\to \infty$, $s\to\infty$. This contradicts the assumption that $c(s)$ has an asymptotic value. Therefore $c(s)\to \infty$ as $s\to \infty$.

\subsubsection*{Zero-beta of the investment opportunity with digital options}

According to the Central Limit Theorem the average of the i.i.d realized payoffs converge in probability to the expected payoff. However, in using the normal approximation for the average payoffs one has to address the problem arising from the fact that for different values of $n$ the distributions of the payoffs from running the bets are different. Therefore one may apply 
a stronger result, namely Berry-Esseen theorem which gives a quantitative control of the convergence speed for the normal approximation of the distribution
in terms of $n$ and the third moment. It suffices to verify that sample correlation between the payoff and the underlying converges to $0$ in probability 
with respect to $\Q$ as $n\to \infty$ because then the same holds for $\P$ also by the equivalence of the measures.

\end{document}